\documentclass[10pt,conference]{IEEEtran}

\usepackage{cite}
\usepackage{amsmath,amssymb,amsfonts}
\usepackage{graphicx}
\usepackage{textcomp}
\usepackage{xcolor}
\usepackage{booktabs}
\usepackage{multirow}

\usepackage{listings}
\usepackage{xcolor}

\usepackage{hyperref}
\hypersetup{
  colorlinks=true,
  linkcolor=blue,
  citecolor=blue,
  urlcolor=blue
}

\definecolor{codebg}{rgb}{0.95,0.98,0.95}

\newcommand{\tbf}[1]{\textbf{#1}}

\begin{document}

\title{HyPulse: A Pulse Synthesis Framework for Hybrid Qubit-Oscillator Gates on Trapped-Ion Platform}

\author{%
  \IEEEauthorblockN{Masoud Hakimi Heris$^{1}$,
                    Yuan Liu$^{1,2,3}$,
                    Frank Mueller$^{2}$ 
                  }
  \IEEEauthorblockA{$^{1}$Department of Electrical and Computer Engineering,
    North Carolina State University, Raleigh, NC 27695, USA}
  \IEEEauthorblockA{$^{2}$Department of Computer Science,
    North Carolina State University, Raleigh, NC 27695, USA}
  \IEEEauthorblockA{$^{3}$Department of Physics,
    North Carolina State University, Raleigh, NC 27695, USA\\
    Corresponding authors: Yuan Liu (q\_yuanliu@ncsu.edu),
    Frank Mueller (fmuelle@ncsu.edu)}
}

\maketitle

\begin{abstract}
  As hybrid qubit-oscillator algorithm development and trapped-ion
  hardware demonstrations advance in parallel, there is a lack of a
  compilation layer connecting the two at the pulse level in the
  vertical software stack. While qubit gate control and pulse
  synthesis are well-established, the translation of hybrid
  qubit-oscillator primitives to the pulse level has not been
  systematically addressed. This gap is further compounded by the
  inherently continuous parametric nature of such gates. Each distinct
  parameter value defines a physically unique operation requiring
  independent pulse optimization, making static pre-compilation
  strategies inapplicable.

  To fill this gap, we present HyPulse, a hardware-aware pulse
  synthesis and generation framework, which contributes a two-phase
  architecture decoupling pulse discovery from circuit assembly. An
  offline optimization engine populates a content-addressed cache of
  high-fidelity primitives: If a pulse for a given gate, parameter,
  and device specification already exists in the library, it is
  retrieved instantly; otherwise the optimizer synthesizes, hashes,
  and caches it automatically. An online assembler then constructs
  circuit-specific pulse programs ready to drive trapped-ion hardware
  control systems via DAX/ARTIQ (Duke) and JaqalPaw/QSCOUT (Sandia),
  trapped-ion pulse execution backends. Beyond circuit execution, this
  architecture enables systematic numerical pre-characterization of
  parametric gate spaces at software timescales before experimental
  deployment.  We demonstrate the framework through the automated
  assembly of a squeezed cat state preparation sequence. By reusing
  cached primitives, HyPulse avoids redundant pulse optimization
  across circuit instances. The results establish HyPulse as a
  hardware-portable pulse synthesis and generation infrastructure for
  hybrid qubit-oscillator trapped-ion processors, with a modular
  design providing the foundation for further developments including
  noise-aware gate scheduling, advanced optimal control techniques,
  and adaptive recalibration.

\end{abstract}

\begin{IEEEkeywords}
Trapped ions, hybrid qubit-oscillator, bosonic qubits, controlled displacement, squeezed cat states, GRAPE, pulse optimization, quantum optimal control, pulse synthesis.
\end{IEEEkeywords}

\section{Introduction}
\label{sec:intro}

Hybrid qubit-oscillator quantum computation encodes information
jointly in qubit and oscillator degrees of freedom, leveraging the
infinite-dimensional Hilbert space of the oscillator to support richer
instruction sets and hardware-efficient error-correcting codes beyond
the reach of qubit-only
architectures~\cite{Andersen2015,Lloyd1999,Weedbrook2012}.  Unlike
purely discrete-variable processors that operate on a finite register
of two-level systems, hybrid processors can exploit the continuous
structure of phase space to encode, manipulate, and protect quantum
information in ways that are fundamentally inaccessible to qubit-only
designs.  This distinction has motivated growing interest in hybrid
qubit-oscillator systems across many applications. Bosonic encodings
such as the Gottesman--Kitaev--Preskill (GKP)
code~\cite{Gottesman2001} and cat qubit
codes~\cite{Mirrahimi2014,Grimsmo2020} encode a logical qubit directly
into an oscillator mode, offering hardware-efficient protection
against dominant decoherence channels, with recent experimental
demonstrations in trapped-ion~\cite{Matsos2025} and
superconducting~\cite{Eickbusch2022} systems confirming their
viability as near-term targets.  In quantum simulation, the native
qubit-oscillator coupling enables direct simulation of vibronic
dynamics~\cite{MacDonell2021,Crane2024HybridFermionBoson} and
spin-boson models~\cite{Montgomery2025HybridSimTI} that are naturally
expressed in a hybrid Hilbert space.  In both domains, the required
operations are precisely controlled qubit-oscillator interactions,
which are the primitives that HyPulse targets.

Recent experimental demonstrations of hybrid oscillator-qubit
simulations~\cite{Montgomery2025HybridSimTI,Crane2024HybridFermionBoson}
and a universal gate set for GKP logical qubits~\cite{Matsos2025}
on trapped-ion hardware have drawn attention to
this platform as a promising substrate for hybrid
qubit-oscillator computation.
The collective motional modes of a trapped-ion chain are native
qumodes that couple directly to the internal qubit states
through laser-driven sideband interactions~\cite{Leibfried2003,
Cirac1995,Haffner2008}, requiring no additional hardware
beyond what is already present in state-of-the-art systems.
Crucially, each ion in the chain couples to all motional modes,
and each motional mode couples to all ions, providing
native all-to-all connectivity between qubit and oscillator
degrees of freedom that is intrinsic to the trapping geometry
rather than engineered through additional circuitry.
This all-to-all structure makes trapped-ion systems a
particularly attractive platform for hybrid qubit-oscillator
computation, where operations between arbitrary qubit-qumode
pairs can be realized without routing overhead.

\begin{figure}[t]
  \centering
  \includegraphics[width=\linewidth]{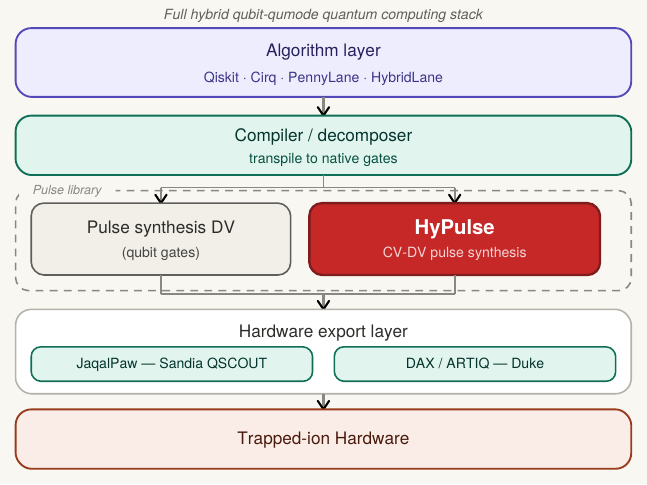}
  \caption{Full hybrid qubit-qumode quantum computing stack.  HyPulse
    fills the hybrid qubit-qumode pulse synthesis slot between the
    compiler/decomposer and the hardware export layer, alongside
    existing DV-only synthesis tools. The content-addressed pulse
    library is shared between the offline synthesis phase and the
    online assembly phase.}
  \label{fig:stack}
\end{figure}

A full quantum computing stack to support such systems, depicted in
Fig.~\ref{fig:stack}, spans four layers: an algorithm layer where
circuits are expressed using high-level gate primitives, a compiler
and decomposer that transpiles to hardware-native gates, a pulse
synthesis layer that converts gate specifications into hardware-ready
waveforms, and a hardware execution layer that drives the physical
control electronics.

Each layer has seen substantial development. At the algorithm layer,
frameworks such as Qiskit, Cirq, and PennyLane support
discrete-variable circuits, while HybridLane~\cite{HybridLane2026}
extends this to hybrid qubit-oscillator instruction
sets~\cite{Liu2026PRX}.  At the hardware execution layer,
ARTIQ/DAX~\cite{ARTIQ, Riesebos2022DAX} and JaqalPaw/QSCOUT~\cite{Jaqal2020} serve
trapped-ion platforms. QICK~\cite{Stefanazzi2022}, designed as a
platform-agnostic open-source qubit controller, and
QuBIC~\cite{Xu2021QuBIC} serve primarily superconducting systems at
the same layer.  At the pulse synthesis layer, tools such as Qiskit
Pulse, PAQOC~\cite{Chen2023PAQOC}, and EPOC~\cite{EPOC2025} address
superconducting DV gates, while Kang et al.~\cite{Kang2021Duke} and
pulselib~\cite{pulselib2024} address trapped-ion DV gates.

The hybrid qubit-oscillator slot of the pulse synthesis layer,
however, has no occupant for trapped-ion platforms.  No existing tool
provides waveform generation for parametric qubit-oscillator gate
primitives on trapped-ion hardware with awareness of sideband coupling
physics and device calibration.
As a result, experimentalists working with hybrid qubit-oscillator
circuits on trapped-ion systems must manually run a separate
gradient-based optimization for each gate and parameter combination,
with no infrastructure for reuse across circuits or sessions, and no
systematic way to validate pulse behavior before committing to
hardware runs.

Beyond the missing stack layer, hybrid qubit-oscillator gates present
a fundamental challenge that has no analogue in the discrete-variable
world: They are inherently parametric.  A controlled displacement
CD($\alpha_1$) and CD($\alpha_2$) for $\alpha_1 \neq \alpha_2$ are
physically distinct operations, each requiring their own pulse
optimization run.  The same holds for controlled squeezing CS($r$) and
controlled rotation CR($\theta$): Every distinct parameter value
defines a unique operation demanding a unique optimized pulse.

In the discrete-variable world this problem does not arise because the
gate set can be finite.  A processor exposes a fixed set of operations
that are pre-compiled once during device calibration and reused across
all circuits.
Even on near-term trapped-ion and superconducting hardware, parametric
single-qubit rotations such as Rz($\theta$) are realized as virtual
phase updates in the control software at zero physical cost for any
angle~\cite{McKay2017}, requiring no pulse optimization.  No analogous
trick exists for hybrid qubit-oscillator primitives, which physically
transform the oscillator state.

This parametric structure creates a concrete bottleneck in
experimental practice.  Without dedicated infrastructure, there is no
mechanism to reuse optimized pulses across circuits or sessions.
Exploring gate behavior over a range of parameter values requires a
separate optimization for each point in the sweep, and the cost
accumulates directly with the number of parameters explored.
As a result, systematic pre-hardware parameter exploration becomes
impractical, and experimentalists are forced to commit to hardware
runs with limited software-level pre-validation.  This is the
fundamental bottleneck that motivates the two-phase architecture of
HyPulse.

We resolve this gap through HyPulse, a two-phase pulse synthesis and
generation framework that decouples the compute-intensive pulse
discovery problem from the latency-critical circuit assembly problem.

\tbf{Phase~1 - Offline Synthesis.}
A gradient-based optimizer such as GRAPE~\cite{Khaneja2005} runs once
per gate primitive and parameter combination.  Optimized pulses are
stored in a content-addressed codebook keyed by a deterministic hash
of the gate specification, device calibration, and Hamiltonian model.
New parameter combinations trigger synthesis on demand and are cached
automatically.

\tbf{Phase~2 - Online Assembler.}
Given a hybrid circuit, the assembler resolves each gate to a codebook
lookup and stitches the waveforms into a hardware-executable pulse
program ready for execution.  Phase~1 is always executed first for the
target gate and parameter combination: If the pulse already exists in
the library it is retrieved instantly; otherwise the optimizer
synthesizes, caches, and returns it automatically.  Therefore, for
circuits composed entirely of previously cached primitives, assembly
is dominated by memory lookup and array concatenation, independent of
circuit depth.  The assembled program is exported directly to
DAX/ARTIQ or JaqalPaw for hardware execution.

The remainder of this paper is organized as follows.
Section~\ref{sec:relatedwork} surveys related pulse-level synthesis
frameworks.
Section~\ref{sec:primitivegates} introduces the three hybrid
qubit-oscillator gate primitives targeted by HyPulse.
Section~\ref{sec:architecture} presents the HyPulse architecture: the
four abstraction layers, content-addressed codebook, device
calibration layer, compile strategies, and an end-to-end example
demonstrating the full pipeline.
Section~\ref{sec:validation} validates all three gate primitives on
published hardware parameters from~\cite{Matsos2025}.
Section~\ref{sec:demo} demonstrates multi-gate circuit assembly
through a squeezed cat state preparation sequence.
Section~\ref{sec:noise} characterizes gate behavior under the
hardware-realistic noise model from~\cite{Matsos2025}.
Section~\ref{sec:discussion} discusses hardware portability and
pre-experimental characterization.
Section~\ref{sec:conclusion} concludes.

\section{Related Work}
\label{sec:relatedwork}

Several prior systems address aspects of pulse-level synthesis for
quantum hardware, but none provides a complete path from hybrid
qubit-oscillator gate primitives to hardware-ready pulses on
trapped-ion systems.

\tbf{Qiskit Pulse}~\cite{Alexander2020QiskitPulse} is a pulse-level
programming framework implemented within Qiskit-Terra that exposes the
OpenPulse interface~\cite{McKay2018OpenPulse} for superconducting
qubit systems.  It allows users to define custom pulse schedules and
perform Hamiltonian characterization and gate calibration at the pulse
level, and has been used to calibrate cross-resonance entangling gates
on IBM Quantum hardware.  Qiskit Pulse operates on a finite DV gate
set for superconducting qubits and has no support for motional modes
or hybrid qubit-oscillator primitives.

\tbf{Boulder Opal}~\cite{BoulderOpal} is a commercial gradient-based
optimal control platform developed by Q-CTRL, providing a flexible
optimization engine for arbitrary quantum systems including
trapped-ion and superconducting hardware.  It supports noise-robust
pulse design and has been used in experimental demonstrations of GKP
logical qubits~\cite{Matsos2025}.  However, Boulder Opal is closed-source and does not provide a built-in content-addressed pulse library tied to device calibration, nor a circuit assembly pipeline targeting trapped-ion control backends such as DAX/ARTIQ and JaqalPaw.

\tbf{PAQOC}~\cite{Chen2023PAQOC} proposes a pulse generation framework
for superconducting DV circuits that mines frequent subcircuits from
quantum programs and converts them into an augmented program-aware
basis gate set, using a criticality-based analytical model to prune
the optimization search space and reduce compilation overhead.  It
operates entirely on superconducting DV gates with no representation
of motional modes or continuously parametric qubit-oscillator gate
specifications.

\tbf{EPOC}~\cite{EPOC2025} proposes an efficient pulse generation
framework for superconducting quantum circuits that combines
ZX-Calculus, circuit partitioning, and circuit synthesis to accelerate
pulse generation and reduce circuit latency by 31.74\% compared to
prior work~\cite{EPOC2025}.  EPOC targets superconducting DV gates and
has no support for motional modes or hybrid qubit-oscillator
primitives.

\tbf{Liang et al.}~\cite{Liang2023HybridGatePulse} propose a hybrid
gate-pulse model for variational quantum algorithms on superconducting
hardware, separating fixed circuit structures from parameterized
layers and applying pulse-level optimization only to the parameterized
portions.  The two-phase separation of optimization cost from circuit
execution is conceptually related to our approach, but their physical
model is entirely discrete-variable with no continuous-variable gate
support.

\tbf{pulselib}~\cite{pulselib2024} provides a graph-based pulse
representation for trapped-ion systems with phase synchronization
capabilities, addressing the hardware representation and scheduling
layer.  It has no optimal control integration and no notion of
individual hybrid qudit-qumode gates, operating below the pulse
synthesis layer rather than within it.

The Duke group's Mølmer--Sørensen pulse-optimization
methods~\cite{Kang2021Duke} apply pulse optimization to two-qubit
gates on trapped-ion hardware. This line of work is the closest to
HyPulse in terms of physics-awareness for trapped-ion systems, but
operates entirely on the qubit subspace with no support for hybrid
qubit--qumode gate primitives.

The common gap across all prior work is clear: No existing open-source
tool provides systematic pulse synthesis for parametric
qubit-oscillator gate primitives on trapped-ion hardware with
device-portable calibration and reusable infrastructure.
Table~\ref{tab:related} summarizes these characteristics.

\begin{table}[h]
\centering
\caption{Comparison of pulse-level synthesis frameworks.
``Duke's MS'' refers to the Mølmer--Sørensen pulse-optimization
methods of Kang et al.~\cite{Kang2021Duke}.}
\label{tab:related}
\setlength{\tabcolsep}{2pt}
\begin{tabular}{lcccccc}
\toprule
\tbf{Feature} & \rotatebox{60}{\tbf{Qiskit Pulse}} &
  \rotatebox{60}{\tbf{Boulder Opal}} &
  \rotatebox{60}{\tbf{PAQOC}} &
  \rotatebox{60}{\tbf{pulselib}} &
  \rotatebox{60}{\tbf{Duke's MS}} &
  \rotatebox{60}{\tbf{HyPulse}} \\
\midrule
Calibration-aware caching     & -- & -- & -- & -- & -- & \checkmark \\
Trapped-ion target    & -- & \checkmark & -- & \checkmark & \checkmark & \checkmark \\
Built-in hybrid gate    & -- & -- & -- & -- & -- & \checkmark \\
Open-source           & \checkmark & -- & \checkmark & \checkmark & -- & \checkmark \\
Hardware export       & \checkmark & \checkmark & -- & \checkmark & -- & \checkmark \\
\bottomrule
\end{tabular}
\end{table}

\section{Hybrid Qubit-Qumode Primitive Gates}
\label{sec:primitivegates}

HyPulse currently targets three fundamental spin--motion gate
primitives drawn from the hybrid oscillator-qubit
ISA~\cite{Liu2026PRX}, each coupling an ancillary spin qubit
to a bosonic motional mode through laser-driven sideband
interactions~\cite{Leibfried2003,Wineland1998}.
The framework is designed to accommodate additional primitives
as the field develops.

\begin{figure}[t]
  \centering
  \includegraphics[width=\linewidth]{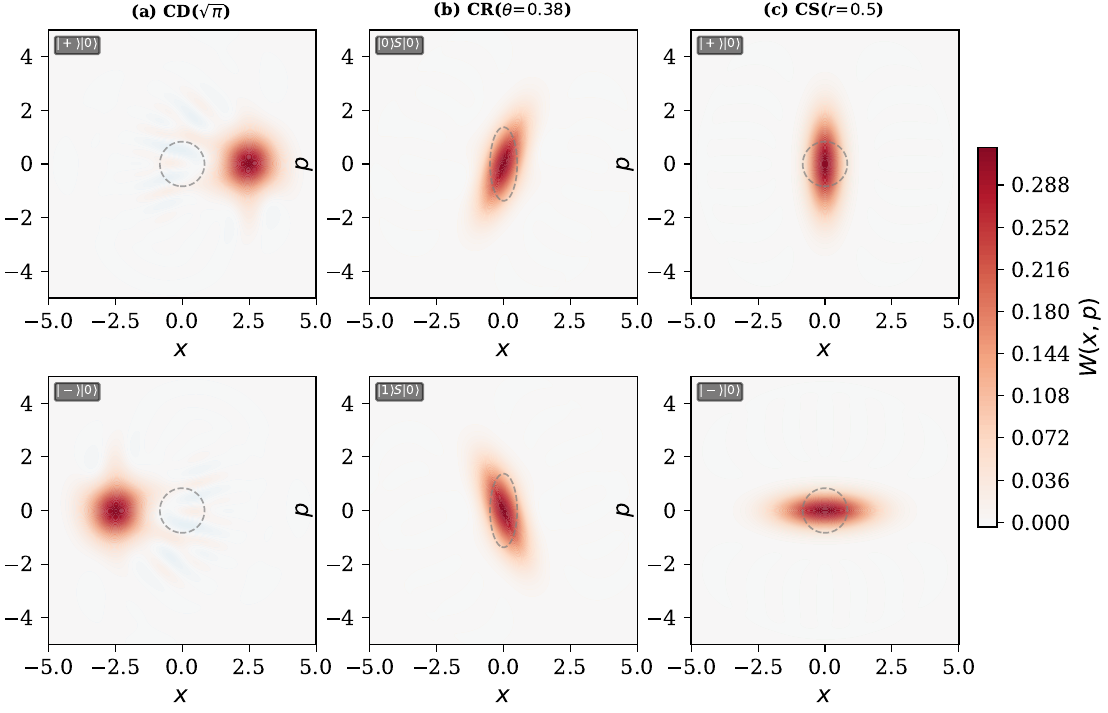}
  \caption{Phase-space illustration of the three HyPulse gate
    primitives. Each column shows one gate. Rows show the
    motional Wigner function conditioned on the two qubit
    branches, with dashed contours showing the initial state.
    (a)~CD($\sqrt{\pi}$): opposite displacements along $x$
    conditioned on $|\pm\rangle$.
    (b)~CR($\theta=0.38$): opposite phase-space rotations of
    a squeezed vacuum conditioned on $|0\rangle$, $|1\rangle$.
    (c)~CS($r=0.5$): opposite squeezings along orthogonal
    quadratures conditioned on $|\pm\rangle$.}  
  \label{fig:gate_primitives}
\end{figure}

\tbf{Controlled Displacement (CD).}
The CD gate is defined as
$U_\mathrm{CD}(\alpha) =
\exp\!\bigl[\sigma_x \otimes (\alpha a^\dagger - \alpha^* a)\bigr]$,
where $\alpha \in \mathbb{C}$ is the displacement amplitude.
In terms of the $\sigma_x$ eigenstates $|\pm\rangle$:
\begin{equation}
  U_\mathrm{CD}(\alpha) =
  |{+}\rangle\langle{+}| \otimes D(+\alpha)
  + |{-}\rangle\langle{-}| \otimes D(-\alpha),
\end{equation}
where $D(\alpha) = e^{\alpha a^\dagger - \alpha^* a}$ is the
displacement operator.
CD applies opposite phase-space displacements conditioned on
the qubit state, and is the elementary operation for cat state
generation~\cite{Eickbusch2022}.

\tbf{Controlled Rotation (CR).}
The CR gate is defined as
$U_\mathrm{CR}(\theta) =
e^{-i\frac{\theta}{2}\,\sigma_z \otimes a^\dagger a}$,
where $\theta$ is the phase-space rotation angle.
In terms of the $\sigma_z$ eigenstates $|0\rangle$, $|1\rangle$:
\begin{equation}
  U_\mathrm{CR}(\theta) =
  |0\rangle\langle 0| \otimes e^{-i\theta a^\dagger a}
  + |1\rangle\langle 1| \otimes e^{+i\theta a^\dagger a}.
\end{equation}
CR applies opposite rotations in phase space conditioned on
the qubit state~\cite{Liu2026PRX}.

\tbf{Controlled Squeezing (CS).}
The CS gate is defined as
$U_\mathrm{CS}(\zeta) =
\exp\!\bigl[\tfrac{1}{2}\,\sigma_x\otimes (\zeta^* a^2 - \zeta a^{\dagger 2})\bigr]$,
where  $\zeta \in \mathbb{C}$  is the squeezing parameter.
In terms of the $\sigma_x$ eigenstates $|\pm\rangle$:
\begin{equation}
  U_\mathrm{CS}(\zeta) =
  |{+}\rangle\langle{+}| \otimes S(+\zeta)
  + |{-}\rangle\langle{-}| \otimes S(-\zeta),
\end{equation}
where $S(\zeta) = \exp\!\bigl[\tfrac{1}{2}(\zeta^* a^2 -
\zeta a^{\dagger 2})\bigr]$ is the single-mode
squeezing operator.
CS drives the second motional sideband and applies opposite
squeezings conditioned on the $\sigma_x$ qubit
eigenstates~\cite{KatzCetinaMonroe2023PRXQ}.

Fig.~\ref{fig:gate_primitives} illustrates the phase-space
effect of each gate, with the two rows showing the motional
Wigner function conditioned on the $|{+}\rangle$ and
$|{-}\rangle$ qubit branches for CD and CS, and the
$|0\rangle$ and $|1\rangle$ branches for CR.
For CD and CS the initial state is the motional vacuum;
for CR a squeezed vacuum is used as the initial state to
make the rotation visually apparent.
In panel (a), CD displaces the motional vacuum in opposite
directions along $x$ depending on the qubit branch, the
hallmark of a spin-conditioned force in phase space.
In panel (b), CR rotates the squeezed state clockwise or
counter-clockwise depending on the qubit branch, turning
the elongated Gaussian by $\pm\theta$ around the origin.
In panel (c), CS squeezes the motional vacuum along
orthogonal quadratures: The $|{+}\rangle$ branch is
squeezed along $p$ while the $|{-}\rangle$ branch is
squeezed along $x$, producing opposite ellipticities
conditioned on the qubit state.

\section{HyPulse Architecture}
\label{sec:architecture}

\begin{figure}[t]
  \centering
  \includegraphics[width=\linewidth]{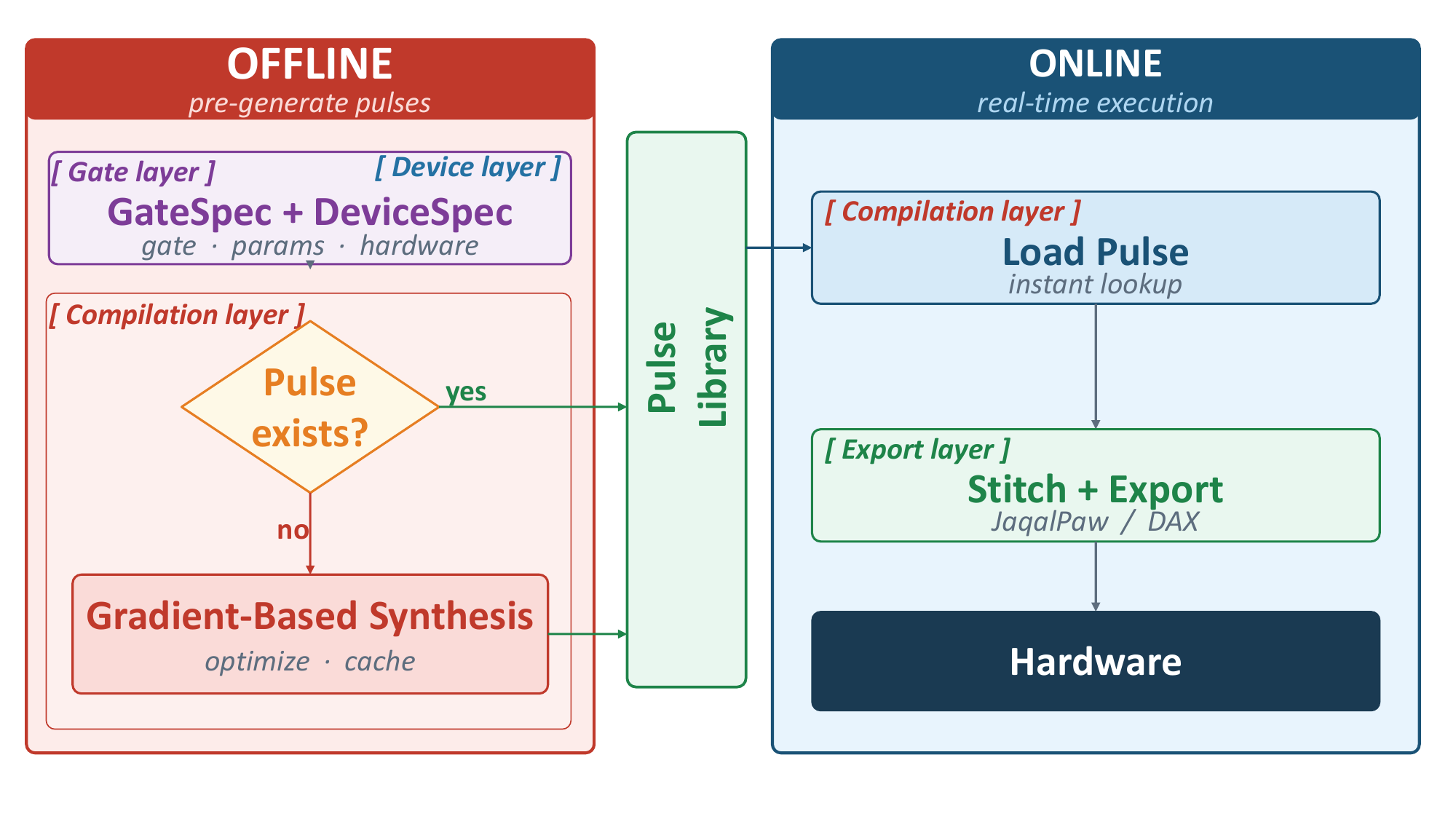}
  \caption{HyPulse two-phase architecture. The offline phase runs
    the optimizer once per (gate, device, model) combination and
    populates a content-addressed library. The online phase assembles
    circuits from cached pulses, triggering
    on-demand synthesis only for parameter combinations not yet
    in the library.}
  \label{fig:workflow}
\end{figure}

\subsection{Core Data Structures and Abstraction Layers}

HyPulse organizes the pulse synthesis pipeline into four
abstraction layers, each hiding complexity from the layers
above it, as illustrated in Fig.~\ref{fig:workflow}.

\tbf{Device layer} abstracts the physical hardware through
\texttt{DeviceSpec}, calibration files, and the
\texttt{CalibrationResolver} protocol.
It encodes the Lamb-Dicke parameter $\eta$, motional mode
frequency $\omega_m$, sideband detuning $\Delta$, maximum drive
amplitude $\Omega_\mathrm{max}$, participation factors $b_{kj}$,
and coherence properties, shielding all upper layers from
hardware-specific physical details.

\tbf{Gate layer} abstracts what to synthesize through
\texttt{GateSpec}, \texttt{ModelSpec}, and the \texttt{GatePlugin}
protocol.
\texttt{GateSpec} encodes the gate type (CD, CR, or CS) and its
continuous parameter, where $\alpha$ is the displacement amplitude
for CD, $\theta$ is the phase-space rotation angle for CR, and $r$
is the squeezing parameter for CS.
\texttt{ModelSpec} encodes the simulation fidelity: Fock space
truncation $n_\mathrm{max}$, time discretization
$N_\mathrm{tslots}$, and optimizer convergence tolerances.
The \texttt{GatePlugin} protocol defines the interface each gate
type must implement: parameter validation and canonicalization,
codebook key construction, Hamiltonian building, and optimizer
invocation.
Together these objects specify \emph{what} to build, entirely
independent of \emph{how} it is built.

\tbf{Compilation layer} implements the two-phase architecture
through \texttt{CompiledGate}, \texttt{compile\_pulse()}, and the
SHA-256 content-addressed library.
\texttt{compile\_pulse()} supports three strategies:
\texttt{lookup} (library only, error on miss),
\texttt{synthesize} (always re-optimize), and
\texttt{hybrid} (default: lookup first, synthesize and cache on
miss).
The \texttt{hybrid} strategy ensures the first call for any
(gate, device, model) combination pays the synthesis cost
once, while all subsequent calls are instant library retrievals.
\texttt{CompiledGate} is the user-facing lazy handle instantiated
via \texttt{CompiledGate.cd}, \texttt{CompiledGate.cr}, and
\texttt{CompiledGate.cs}: Calling \texttt{cg.pulse()} triggers
the full pipeline on demand, and calling \texttt{cg.unitary()}
additionally propagates the waveform through the Hamiltonian to
compute the process unitary.

\tbf{Export layer} translates cached pulses into hardware-ready
schedules through \texttt{PulseSpec}, \texttt{pulse\_concat},
\texttt{CircuitExporter}, and the DAX/ARTIQ and JaqalPaw backends.
It reconstructs the physical laser frequency
$\omega_L = \omega_q - \omega_m + \Delta$ from stored device
parameters, where $\omega_q$ is the qubit transition frequency,
$\omega_m$ is the motional mode frequency, and $\Delta$ is the
sideband detuning, and assembles the full pulse sequence with
appropriate inter-gate timing buffers.
A typical workflow consists of three steps: construct a
\texttt{CircuitExporter} with the target device and model,
add gates via \texttt{add\_cs()}, \texttt{add\_cd()}, or
\texttt{add\_cr()}, and call \texttt{export()} to produce
hardware-ready pulse schedules for both DAX/ARTIQ and JaqalPaw.

The four layers interact strictly through their interfaces: a user
touches the gate layer (via \texttt{CompiledGate.cd} and
similar) and the export layer (via \texttt{CircuitExporter}).
The device and compilation layers are internal to the framework.  
Table~\ref{tab:layers} summarizes the key components of each layer.

\begin{table}[h]
\centering
\caption{HyPulse four abstraction layers and their key components.}
\label{tab:layers}
\footnotesize
\setlength{\tabcolsep}{2pt}
\begin{tabular}{lp{3cm}p{3cm}}
\toprule
\tbf{Layer} & \tbf{Key Components} & \tbf{Role} \\
\midrule
Device      & \texttt{DeviceSpec}, \texttt{CalibResolver}
            & Hardware parameters, mode participation \\
Gate        & \texttt{GateSpec}, \texttt{ModelSpec}, \texttt{GatePlugin}
            & Synthesis specification \\
Compilation & \texttt{CompiledGate}, \texttt{compile\_pulse()}, SHA-256
            & Two-phase caching and strategies \\
Export      & \texttt{CircuitExporter}, DAX/JaqalPaw
            & Hardware-ready pulse schedules \\
\bottomrule
\end{tabular}
\end{table}

\subsection{Content-Addressed Codebook}
\label{subsec:codebook}
The codebook key is computed as
\texttt{k = SHA256(serialize(G, D, M))},
where \texttt{G}, \texttt{D}, and \texttt{M} are the
canonicalized \texttt{GateSpec}, \texttt{DeviceSpec},
and \texttt{ModelSpec}, respectively.
Canonicalization ensures deterministic hashing regardless of
floating-point representation: All floats are serialized to 17 significant digits to guarantee IEEE~754 double-precision round-trip equality; dictionary keys are sorted lexicographically to produce deterministic JSON regardless of insertion order; complex parameters such as the displacement amplitude $\alpha \in \mathbb{C}$ are decomposed into ordered \texttt{(real, imaginary)} tuples to avoid representation ambiguity; and non-finite values (\texttt{NaN}, $\pm\infty$) raise an error rather than producing a silently incorrect hash. Together, these rules ensure that two physically identical specifications always produce the same key on any machine, while any physically meaningful change, however small, produces a different key.
This design has two important properties beyond performance
caching.
First, it is \emph{collision-resistant}: Two physically distinct
gate instances always map to different keys.
Second, it is a \emph{physical validity certificate}: If any
element of the triple changes, including device calibration drift
in $\eta$, a shift in $\omega_m$, or a recalibration of
$\Omega_\mathrm{max}$, the key changes and the stale cached pulse
is automatically invalidated without any explicit cache management.

For example, CD($\sqrt{\pi}$) and CD($2\sqrt{\pi}$) on the same
device and model produce distinct keys:
\begin{lstlisting}
k1 = _request_hash_for_library(GateSpec("CD", {"alpha_re": 1.7724, ...}),
                  device, model)
# k1: "3a7f2c..."

k2 = _request_hash_for_library(GateSpec("CD", {"alpha_re": 3.5449, ...}),
                  device, model)
# k2: "9b1e4d..."  -- distinct, no collision
\end{lstlisting}
A recalibration that updates $\eta$ from $0.083$ to $0.085$
produces a new key for every gate on that device, automatically
invalidating all stale cached pulses without any explicit cache
management logic.

\subsection{Compile Strategies}

The central function of the HyPulse pipeline is
\texttt{compile\_pulse(gate, device, model, strategy)}, which
implements the two-phase architecture through three selectable
strategies:

\tbf{lookup}: Search the library for an existing pulse matching
the (gate, device, model) triple. Return immediately on a hit;
raise an error on a miss. Used when synthesis is not permitted
at runtime.

\tbf{synthesize}: Always run the optimizer, regardless of whether
a cached pulse exists. Used for explicit re-synthesis or
benchmarking.

\tbf{hybrid} (default): Attempt a library lookup first. On a
cache hit, return the stored \texttt{PulseSpec} immediately.
On a miss, run the optimizer, hash and save the result to the
library under its deterministic path, and return the new pulse.
The next call with identical (gate, device, model) finds the
stored pulse and skips synthesis entirely.

The \texttt{hybrid} strategy is HyPulse's two-phase architecture:
The first call for any parameter combination pays the synthesis cost;
all subsequent calls are free.

\subsection{User-Facing Interface: CompiledGate}

Users interact with HyPulse through the \texttt{CompiledGate}
object, instantiated via \texttt{CompiledGate.cd},
\texttt{CompiledGate.cr}, and \texttt{CompiledGate.cs}.
It stores the gate, device, and model specifications and
defers all computation until explicitly requested.
On lookup, it first attempts an exact hash match; if not
found, it retrieves the highest-fidelity cached pulse for
the same gate and device, falling back to synthesis only
when no cached pulse exists.
This fuzzy fallback enables a common workflow in which Phase~1
synthesizes a pulse with explicit optimizer parameters
(\texttt{duration\_us}, \texttt{num\_tslots}, \texttt{amp\_bound},
etc.), while Phase~2 retrieves the cached pulse using only the
gate's physics parameters---without needing to repeat the
optimizer settings.

Calling \texttt{cg.pulse()} triggers the full pipeline,
including library lookup and synthesis on a miss, and returns
the \texttt{PulseSpec}.
Calling \texttt{cg.unitary()} first calls \texttt{cg.pulse()},
then propagates the optimized waveform through the spin-motion
Hamiltonian to obtain the process unitary, caching the result
separately for subsequent calls.

This lazy evaluation model means that constructing a circuit
of $N$ gates incurs no synthesis cost until \texttt{pulse()} or
\texttt{unitary()} is called, and only gates not already in the
library trigger new optimization.

\subsection{Device Calibration Layer}

HyPulse includes a calibration layer that maps hardware-specific
calibration data to the effective Lamb-Dicke parameter $\eta$
consumed by the synthesis layer.
In a trapped-ion chain, each ion couples to each collective
motional mode with a different strength, captured by the
mode participation factor $b_{kj} = \eta_{kj}/\eta_k$,
where $\eta_k$ is the single-ion Lamb-Dicke parameter for
mode $k$ and $b_{kj}$ is the normalized participation of
ion $j$ in that mode~\cite{Leibfried2003,Sun2025SpinBoson}.
Calibration drift between experimental sessions is handled by the same content-addressed mechanism described in Sec.~\ref{subsec:codebook}. Consider a workflow in which an experimentalist begins a session with a recalibrated device specification \texttt{sydney\_gkp\_2026\_04\_27} reflecting the latest measured $\eta$, $\omega_m$, and $\Omega_{\max}$. Each call to \texttt{CompiledGate.cd}, \texttt{cr}, or \texttt{cs} computes a key over this updated \texttt{DeviceSpec}, so any pulse cached under the previous calibration produces a different hash and is automatically bypassed. The optimizer is invoked on demand to populate the new entries, while pulses from prior sessions remain on disk---never served, never deleted---until garbage collection if desired. The user writes no cache-management code; physical validity is enforced by the key construction rather than by runtime checks.

\subsection{Modular Extensibility}

HyPulse extends along two independent axes, each requiring
minimal new code.

\tbf{New gate primitives.}
Adding a gate type requires implementing the \texttt{GatePlugin}
protocol, which consists of four methods: \texttt{validate\_and\_normalize}
(parameter validation and canonicalization), \texttt{pulse\_lookup\_key}
(codebook key construction), \texttt{synthesize} (optimizer
invocation), and \texttt{build\_hamiltonian} (spin-motion
Hamiltonian construction).
Once registered via \texttt{register(plugin)}, the new primitive
is automatically supported by the codebook, online assembler,
and hardware export layer without any changes to those components.

\begin{lstlisting}[language=Python,caption={Minimal \texttt{GatePlugin} skeleton for a hypothetical controlled-parity gate.},label={lst:plugin},basicstyle=\ttfamily\footnotesize,frame=single]
class CParityPlugin(GatePlugin):
    gate_type = "CParity"

    def validate_and_normalize(self, params):
        # canonicalize phi to [0, 2*pi)
        return {"phi": float(params["phi"]) % (2*math.pi)}

    def pulse_lookup_key(self, gate, device, model):
        return canonical_hash(gate, device, model)

    def build_hamiltonian(self, gate, device, model):
        # construct spin-motion H for parity interaction
        return H_drift, [H_ctrl_x, H_ctrl_p]

    def synthesize(self, gate, device, model):
        return run_grape(self.build_hamiltonian(...),
                         target=parity_unitary(gate.phi))

register(CParityPlugin())
\end{lstlisting}

Listing~\ref{lst:plugin} sketches a minimal plugin definition. The plugin author specifies how parameters are validated, how the codebook key is constructed, and how the spin--motion Hamiltonian is assembled; all caching, scheduling, and hardware export behavior is inherited from the framework.

\tbf{New hardware targets.}
Adding a new trapped-ion platform requires only a calibration
file specifying $\eta$, $\omega_m$, $\Delta$,
$\Omega_\mathrm{max}$, and coherence properties.
If the calibration format matches an existing resolver, no new
code is required.
If the format is novel, a single resolver class implementing the
\texttt{CalibrationResolver} protocol is sufficient, and all
other framework components remain unchanged.
The resolver exposes a uniform interface to the synthesis layer:
given an ion-mode pair, it returns the effective Lamb-Dicke
parameter $\eta_{kj}$ along with any device-specific extras,
while the remaining hardware quantities ($\Omega_\mathrm{max}$,
sideband detuning, coherence properties) are read directly from
the corresponding \texttt{DeviceSpec}.
This separation enables a single pulse library to span multiple
devices---each indexed under its own \texttt{DeviceSpec}---without
risk of cross-contamination, since the content-addressed key
distinguishes pulses by device identity. In practice, retargeting
HyPulse from one trapped-ion system to another reduces to writing
a small Python calibration module (a few dozen lines defining the
\texttt{DeviceSpec} and resolver bindings); the gate plugins,
optimizer, codebook, and export backends are reused unchanged.

The two axes are orthogonal: A new gate primitive works on all
existing hardware targets, and a new hardware target supports
all existing gate primitives.

\subsection{Offline Phase: Pulse Synthesis}

For each (GateSpec, DeviceSpec, ModelSpec) triple not present in
the codebook, HyPulse constructs the appropriate spin-motion
Hamiltonian via the registered \texttt{GatePlugin} and runs
gradient-based optimal control via qutip-qtrl~\cite{QuTiP,Khaneja2005}.
Amplitude bounds are derived from the device calibration,
ensuring the optimized controls remain within hardware limits.
The synthesized \texttt{PulseSpec} is stored under its
deterministic hash path (SHA-256) along with all physical metadata
required for hardware export.

\subsection{Online Phase: Circuit Assembly}

Given a circuit expressed as a sequence of \texttt{CompiledGate}
objects, the online assembler:
\begin{enumerate}
  \item Calls \texttt{cg.pulse()} for each gate.
        Cache hits return immediately; cache misses trigger
        on-demand synthesis, storage, and retrieval via the
        \texttt{hybrid} strategy.
  \item Concatenates piecewise-constant waveforms with
        appropriate inter-gate timing buffers.
  \item Exports the assembled pulse program to DAX/ARTIQ or
        JaqalPaw, reconstructing the physical laser frequency
        $\omega_L = \omega_q - \omega_m + \Delta$ from the
        stored device parameters.
\end{enumerate}
For circuits composed entirely of cached primitives, assembly
time is dominated by memory lookup and array concatenation,
achieving performance independent of gate count.

\subsection{End-to-End Example}

The following listings illustrate the complete HyPulse pipeline
for a CD($0.5+0.5i$) gate on the hardware
parameters reported by Matsos et al.~\cite{Matsos2025},
demonstrating all four abstraction layers in sequence.
Table~\ref{tab:sydney_params} summarizes the device parameters
used throughout this work.
In Phase~1, the gate is synthesized for the first time and
cached in the content-addressed library.
In Phase~2, the same call returns instantly from cache, the
pulse unitary is evaluated, and the circuit is exported to
both DAX/ARTIQ and JaqalPaw hardware backends.

\tbf{Phase~1 --- Offline Synthesis.}
\begin{lstlisting}
from hypulse.calibrations import get_device_spec
from hypulse.core import CompiledGate
import numpy as np

device = get_device_spec("sydney_gkp_v1")

# Explicit physics parameters for first-time synthesis
cg = CompiledGate.cd(0.5+0.5j,
                     device=device,
                     qubit=0, qumode=0,
                     cutoff=15,
                     duration_us=100.0,
                     num_tslots=200,
                     amp_bound=1.0,
                     Delta_rad_per_s=2*np.pi*10e3,
                     match="exact")
pulse = cg.pulse()  # synthesizes, hashes, saves
\end{lstlisting}

\tbf{Phase~2 --- Online Assembly.}
\begin{lstlisting}
# Cache retrieval
cg = CompiledGate.cd(0.5+0.5j,
                     device=device,
                     qubit=0, qumode=0,
                     cutoff=15)
pulse = cg.pulse()   # cache hit

# Evaluate
U = cg.unitary()
F = process_fidelity(U_target, U)
\end{lstlisting}

\tbf{Phase~2 --- Hardware Export.}
\begin{lstlisting}
# Circuit assembly and hardware export
exporter = CircuitExporter(device, ModelSpec(cutoff=15))
exporter.add_cd(0.5+0.5j)
result = exporter.export("cd_demo",
                         repo_root=Path("."))
# Exports DAX/ARTIQ and JaqalPaw schedules
\end{lstlisting}

\begin{table}[h]
\centering
\caption{Hardware parameters derived from Matsos et al.~\cite{Matsos2025}, corresponding to the radial-$x$ motional mode.}
\label{tab:sydney_params}
\setlength{\tabcolsep}{5pt}
\begin{tabular}{lll}
\toprule
\tbf{Parameter} & \tbf{Value} & \tbf{Description} \\
\midrule
$\eta$           & $0.083$                & Lamb-Dicke parameter \\
$\omega_m$       & $2\pi \times 1.33$~MHz & Motional mode frequency \\
$\Omega_j$       & $2\pi \times 2.4$~kHz  & Sideband Rabi rate \\
\bottomrule
\end{tabular}
\end{table}

\section{Gate Primitives Validation}
\label{sec:validation}

We validate all three gate primitives using hardware parameters reported by Matsos et al. ~\cite{Matsos2025}:
$\eta = 0.083$, $\omega_m = 2\pi \times 1.33$~MHz, and
$\Omega_j = 2\pi \times 2.4$~kHz. Fock space truncation
$n_\mathrm{max} = 15$ is used throughout.
The validation parameters are chosen to exercise each primitive
at a physically meaningful operating point: $\alpha = \sqrt{\pi}$
for CD corresponds to the unit cell of the GKP code lattice and
the elementary displacement used in cat-state generation; 
$\theta = 0.382$ for CR is a representative phase-space rotation
angle within the linear regime of dispersive coupling; and
$r = 0.5$ for CS produces approximately $4.3$~dB of squeezing,
sufficient to clearly resolve quadrature asymmetry while remaining
within the second-sideband coupling regime.

Table~\ref{tab:gate_fidelities} confirms that all three primitives
achieve closed-system fidelity above 0.99 on the mentioned hardware
parameters above. CD and CR reach target fidelity within 100--300~\textmu s;
CS requires a longer duration of 1300~\textmu s owing to its
second-sideband coupling, which accumulates squeezing more slowly
than the first-sideband displacement and rotation operations.

\begin{figure}[t]
  \centering
  \includegraphics[width=\linewidth]{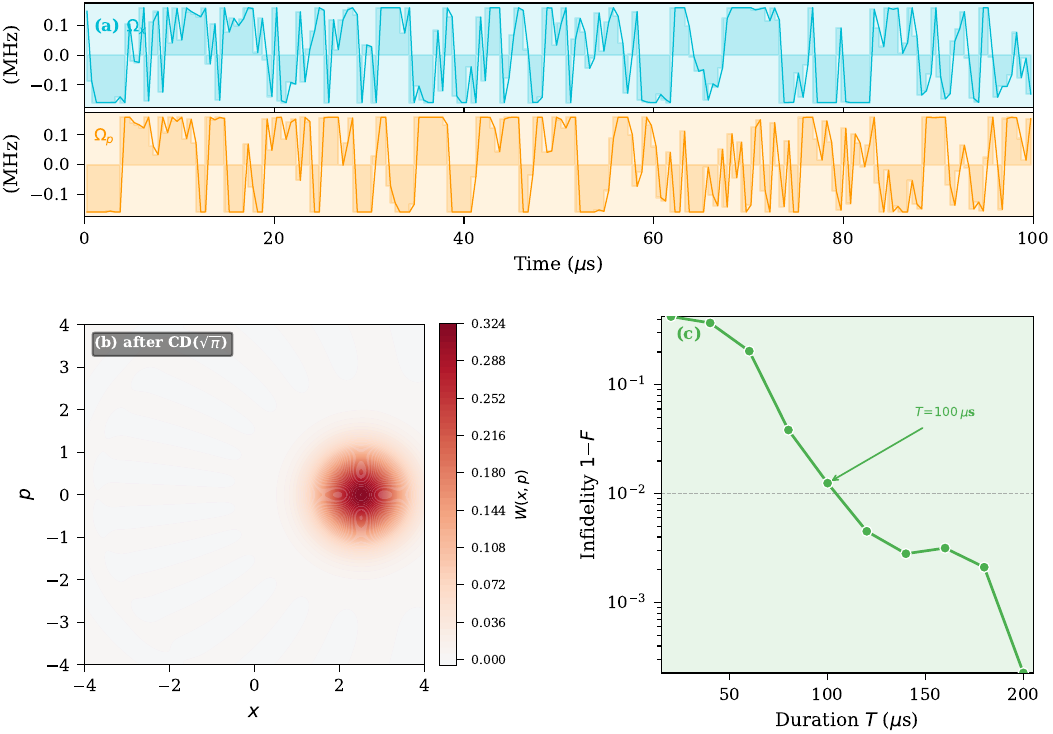}
  \caption{CD gate characterization on the hardware parameters from Table \ref{tab:sydney_params}.
    (a) Optimized pulse waveforms $\Omega_x(t)$ (blue) and
    $\Omega_p(t)$ (orange) for $T = 100$~\textmu s, showing
    both channels bounded within the hardware amplitude limit.
    (b) Wigner function of the motional mode after CD($\sqrt{\pi}$),
    confirming a coherent displacement.
    (c) Infidelity $1 - \mathcal{F}$ vs.\ gate duration $T$,
    demonstrating the pre-characterization sweep capability
    of the framework.}
  \label{fig:cd_sweep}
\end{figure}

Fig.~\ref{fig:cd_sweep} shows the CD gate in detail.
Panel (a) presents the optimized pulse waveforms $\Omega_x(t)$
and $\Omega_p(t)$ for $T = 100$~\textmu s, showing the two control
channels.
Panel (b) shows the Wigner function of the motional mode after
applying CD($\sqrt{\pi}$) to the initial state
$|{+}\rangle \otimes |0\rangle$, confirming the expected single
coherent lobe at displacement $\alpha = \sqrt{\pi}$.
Panel (c) shows the infidelity $1 - \mathcal{F}$ as a function of
gate duration $T$: The operating point $T = 100$~\textmu s lies
just below the $10^{-2}$ infidelity threshold, and longer durations
yield further improvement.
This duration-fidelity sweep illustrates the pre-hardware
characterization capability of HyPulse: Experimentalists can
explore the full parameter space in software before committing
hardware time.

\begin{table}[h]
\centering
\caption{Gate primitive closed-system fidelities on Sydney
  $^{171}$Yb$^+$ hardware parameters ($\eta = 0.083$,
  $n_\mathrm{max} = 15$). All gates achieve $\mathcal{F} \geq 0.99$.}
\label{tab:gate_fidelities}
\setlength{\tabcolsep}{5pt}
\begin{tabular}{lccc}
\toprule
\tbf{Gate} & \tbf{Parameter} & \tbf{Duration} & $\mathcal{F}$ \\
\midrule
CD & $\alpha = \sqrt{\pi}$ & 100~\textmu s  & $0.994$ \\
CR & $\theta = 0.382$      & 300~\textmu s  & $0.996$ \\
CS & $r = 0.5$             & 1300~\textmu s & $0.99999$ \\
\bottomrule
\end{tabular}
\end{table}

\section{Squeezed Cat State Demonstration}
\label{sec:demo}

To demonstrate multi-gate circuit assembly, we prepare a squeezed
cat state, a non-classical superposition of two displaced squeezed
states whose Wigner function exhibits both spatial separation and
quantum interference fringes~\cite{Kienzler2015}.

\subsection{Circuit}

The preparation circuit consists of three operations:
\begin{enumerate}
\item CS($r = 0.5$) squeezes the motional vacuum along the $X$
  quadrature, creating an elongated Gaussian with reduced quantum
  noise in one direction.
\item This is followed by a single-qubit Hadamard on the spin,
  rotating from the $\sigma_x$ to the $\sigma_z$ eigenbasis.
\item Finally, CD($\alpha = \sqrt{\pi}$) applies a conditional
  displacement, entangling the spin and the motional mode.
\end{enumerate}
Post-selection on the spin state $|\uparrow\rangle$ then projects the
motional mode onto the even squeezed cat state:
\begin{equation}
  |\psi_\mathrm{cat}\rangle \propto
  \bigl(D(\sqrt{\pi}) + D(-\sqrt{\pi})\bigr) S(0.5)|0\rangle.
\end{equation}
Both primitives are retrieved from the codebook; no optimization
runs at assembly time.
This is the first multi-gate circuit assembled by HyPulse and exercises
the full pipeline end-to-end: two distinct gate plugins (CS and CD)
contribute pulses synthesized in separate Phase~1 calls, the online
assembler concatenates their waveforms with appropriate inter-gate
buffers, and the resulting program is exported through the same
DAX/ARTIQ and JaqalPaw backends used for single-gate validation.
A successful preparation therefore tests not only the individual
primitives but also the codebook lookup, waveform stitching, and
schedule export logic in a single experiment.

\subsection{Results}

\begin{figure}[t]
  \centering
  \includegraphics[width=\linewidth]{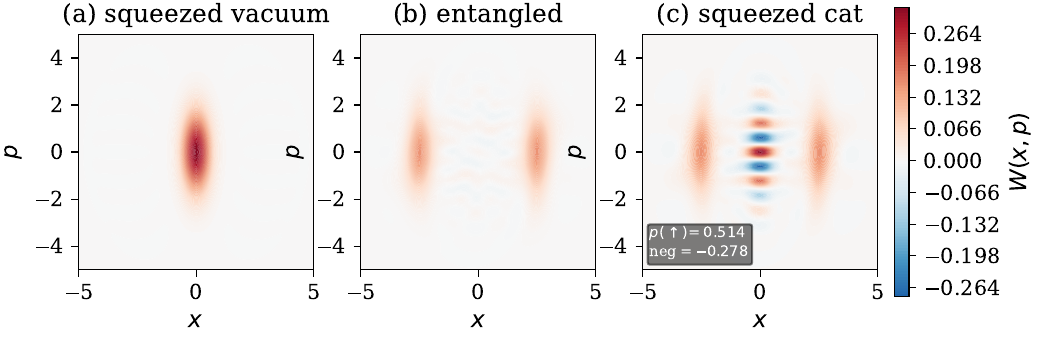}
  \caption{Wigner function evolution through the squeezed cat
    preparation circuit on Sydney hardware parameters.
    (a) After CS($r = 0.5$): squeezed vacuum.
    (b) After CS + CD($\sqrt{\pi}$): spin-motion entangled state
    with two separated lobes.
    (c) After post-selection on $|\uparrow\rangle$: squeezed cat
    state with Wigner negativity $\mathrm{neg} = -0.278$,
    confirming non-classical coherence consistent with
    Kienzler et al.~\cite{Kienzler2015}.}
  \label{fig:squeezed_cat}
\end{figure}

Fig.~\ref{fig:squeezed_cat} shows the Wigner function at each circuit
stage.
After CS (panel a), the vacuum Gaussian is elongated along $P$,
confirming squeezing.
After CD (panel b), the state is entangled: Two separated squeezed
lobes are visible, corresponding to the two spin-conditioned
displacement branches.
After post-selection on $|\uparrow\rangle$ (panel c), the motional
mode is projected onto the squeezed cat state with peak Wigner
negativity $\mathrm{neg} = -0.278$ and post-selection probability
$p(\uparrow) = 0.514 \approx 0.5$, consistent with the expected
even-parity cat state.

The clear Wigner negativity in panel (c) is the definitive signature
of non-classical coherence, confirming that the two HyPulse-compiled
primitives compose correctly without introducing phase errors or
timing artifacts at the stitch boundary.

\section{Noise Characterization}
\label{sec:noise}

A primary use case of HyPulse is systematic numerical
pre-characterization of gate behavior under hardware-realistic
noise before experimental deployment.
We simulate each primitive under the motional dephasing
model reported by Matsos et al.~\cite{Matsos2025}:
The noisy Hamiltonian includes a dephasing term
$H_\mathrm{deph}(t) = \epsilon(t)\, a^\dagger a$,
where $\epsilon(t)$ is a piecewise-constant Gaussian process
with variance $\sigma^2 = 2\gamma/\tau_\mathrm{seg}$,
decay rate $\gamma = 18$~Hz, and segment duration
$\tau_\mathrm{seg}$ matching the pulse time discretization.
Results are averaged over 500 independent noise trajectories.

\begin{table}[h]
\centering
\caption{Gate fidelities under Sydney's motional dephasing
  model ($\gamma = 18$~Hz, 500 trajectories).}
\label{tab:noise}
\setlength{\tabcolsep}{4pt}
\begin{tabular}{lcccc}
\toprule
\tbf{Gate} & \tbf{Duration} & $\mathcal{F}_\mathrm{closed}$
  & $\mathcal{F}_\mathrm{noisy}$ & $\Delta\mathcal{F}$ \\
\midrule
CD & 100~\textmu s  & $0.994$ & $0.922 \pm 0.04$ & $0.07$ \\
CR & 300~\textmu s  & $0.996$ & $0.850 \pm 0.18$ & $0.15$ \\
CS & 1300~\textmu s & $0.99999$ & $0.554 \pm 0.33$ & $0.45$ \\
\bottomrule
\end{tabular}
\end{table}

\begin{figure}[t]
  \centering
  \includegraphics[width=\linewidth]{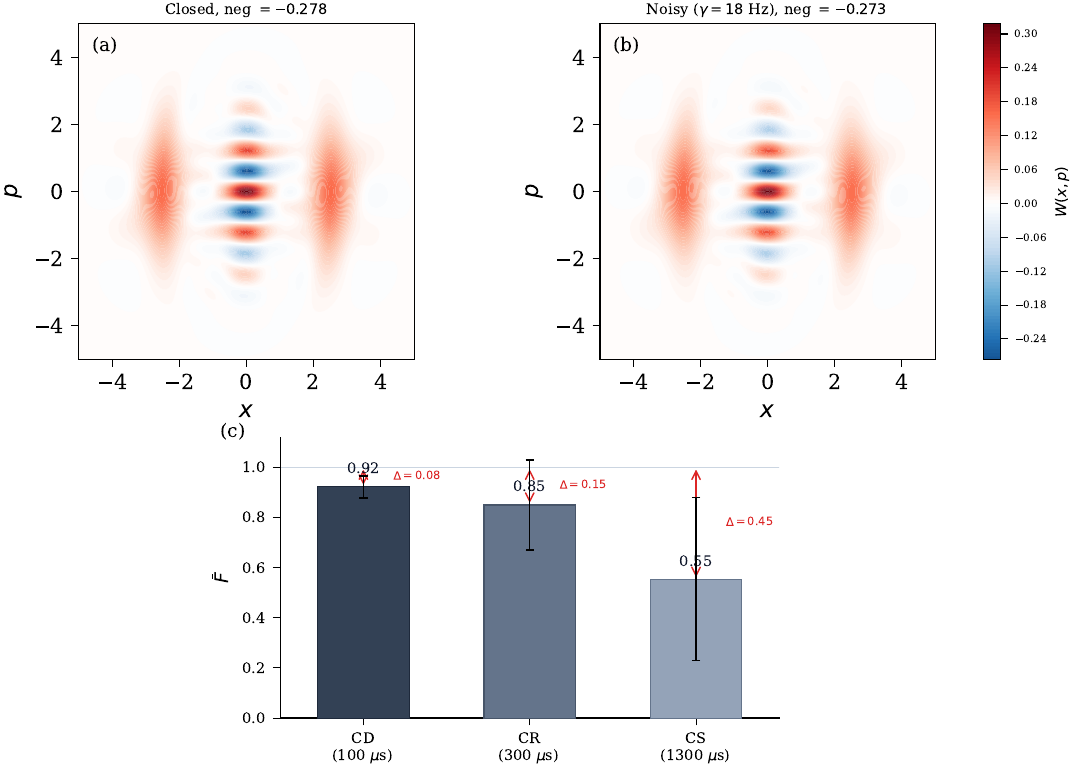}
  \caption{Noise characterization under the motional
    dephasing model reported in Matsos et al.\cite{Matsos2025} ($\gamma = 18$~Hz, 500 trajectories).
    (a) Closed-system squeezed cat Wigner function,
    $\mathrm{neg} = -0.278$.
    (b) Noisy Wigner function averaged over 500 trajectories,
    $\mathrm{neg} = -0.273$: negativity preserved.
    (c) Per-gate noisy fidelity $\mathcal{F}_\mathrm{noisy}$
    for CD, CR, and CS.
    CD and CR are less affected by noise; CS is limited by its
    1300~\textmu s gate duration, a hardware constraint
    addressable by improved sideband Rabi rates.}
  \label{fig:noise}
\end{figure}

Table~\ref{tab:noise} and Fig.~\ref{fig:noise} summarize
the results.
CD and CR are less affected by noise at current
parameters, with fidelity degradation of 7\% and 15\%,
respectively, consistent with their gate durations of
100~\textmu s and 300~\textmu s under the
$\gamma = 18$~Hz dephasing rate.
CS is noise-limited: Its second-sideband coupling requires
1300~\textmu s to accumulate $r = 0.5$ squeezing, and
dephasing over this duration introduces $\Delta\mathcal{F}
\approx 0.45$ with high variance, indicating sensitivity
to the noise realization.
This reflects current hardware constraints, not an
architectural limitation of HyPulse.
Improved sideband Rabi rates on future hardware would
directly reduce the CS gate time and suppress the noise
floor.
Identifying this constraint in software, before any trap
time is committed, illustrates HyPulse's role as a
pre-experimental screen: the framework makes it possible to
quantify the gap between current and required hardware
performance ahead of deployment, and to translate noise
budgets directly into hardware development targets such as
required sideband Rabi rates and motional coherence times.

Crucially, the squeezed cat circuit preserves Wigner
negativity under noise.
The peak negativity degrades only from $-0.278$ to
$-0.273$, a 1.8\% reduction that leaves the interference
fringes clearly visible.
The post-selection on $|\uparrow\rangle$ projects the motional
mode onto the cat superposition, and the modest per-lobe
squeezing reduction under dephasing does not destroy the
interference structure responsible for Wigner negativity.

\section{Discussion}
\label{sec:discussion}

\tbf{Physical validity as a structural property.}
The content-addressed key makes physical validity a structural
property of the cache rather than a runtime check.
A pulse synthesized for $\eta = 0.083$ is never served to a circuit
targeting $\eta = 0.10$; and a pulse cached before a trap recalibration
is never reused after.
This property is especially important for long-running experimental
campaigns where device parameters drift between sessions, and removes
an entire class of subtle correctness bugs from the user's
responsibility.

\tbf{Hardware portability.}
Retargeting HyPulse to a new trapped-ion platform requires only a
new calibration file.
The framework supports the hardware reported in \cite{Matsos2025}, producing distinct pulse
libraries that correctly reflect the different physical coupling
strengths and mode structures.

\tbf{Pre-experimental characterization as a first-class use case.}
The duration-fidelity sweep of Fig.~\ref{fig:cd_sweep}(c) and the
noise characterization of Sec.~\ref{sec:noise} illustrate HyPulse's
role as a pre-experimental exploration environment.
Experimentalists can identify viable operating points, assess noise
sensitivity, and project the impact of hardware improvements entirely
in software before committing trap time.
The CS noise results in particular quantify how improved sideband
Rabi rates would reduce gate duration and suppress the noise floor,
providing a concrete software-derived target for hardware development.

\tbf{Limitations and threats to validity.}
The validation in this paper is entirely numerical: pulses are
synthesized and characterized in simulation against the device
parameters reported in~\cite{Matsos2025}, and the noise model is
restricted to the dominant motional dephasing channel.
Hardware-specific effects such as cross-mode coupling, AC Stark
shifts, laser intensity noise, and finite spin coherence are not
included in the present analysis and would shift the achievable
fidelities in practice.
Validation against live hardware data, integration with additional
trapped-ion platforms beyond the reference parameters, and extension
of the noise model to incorporate vendor-supplied noise spectra are
the natural next steps and are scoped as future work in
Sec.~\ref{sec:future}.
\section{Conclusion}
\label{sec:conclusion}

We presented \tbf{HyPulse}, an open-source two-phase pulse synthesis
framework that fills the gap in the vertical quantum stack of hybrid
qubit-oscillator trapped-ion architectures. The framework addresses
the fundamental challenge of continuously parametric gates: By
separating the one-time offline synthesis cost from repeated online
circuit assembly, HyPulse provides the infrastructure for systematic
pulse reuse, hardware-portable calibration, and pre-experimental
parameter exploration that is currently absent for hybrid
qubit-oscillator systems.

Validated on the hardware parameters from \cite{Matsos2025}, HyPulse
achieves closed-system fidelity above 0.99 for all three primitives
and assembles a squeezed cat state preparation circuit with
preserved Wigner negativity ($\mathrm{neg} = -0.273$) under
the reported dephasing model in \cite{Matsos2025}.
CD and CR are less affected by noise at current hardware parameters.
CS noise sensitivity is tied to its longer gate duration, a
hardware constraint that improved sideband Rabi rates would
directly address.

The modular architecture of HyPulse is designed to grow with
the field.
As new hybrid gate primitives are proposed and new trapped-ion
platforms come online, the framework absorbs them through its
plugin and calibration interfaces.
As optimization techniques mature, they plug directly into the
offline synthesis phase without affecting the rest of the stack.

Together, these properties position HyPulse as a foundation for
systematic, reusable, and hardware-aware pulse synthesis in the
hybrid qubit-oscillator stack, complementing existing
discrete-variable tooling and lowering the barrier to
experimental exploration of bosonic quantum information
processing on trapped-ion devices.
\section{Future works}
\label{sec:future}

Future directions include noise-aware pulse synthesis that
incorporates hardware noise spectra directly into the
optimization objective, multi-gate joint optimization that
exploits pulse overlap between adjacent gates, and extension
of the gate primitive set to include two-qubit hybrid
entangling operations and controlled parity gates.
On the hardware side, integration with additional trapped-ion
platforms and validation against experimental data from
live hardware runs remain important next steps.

Beyond these, several directions follow naturally from the
two-phase architecture. Adaptive recalibration could close the
loop between cached pulses and live device feedback, automatically
re-synthesizing primitives when measured fidelity drops below a
threshold. The content-addressed library is also a natural target
for transfer-learning approaches, in which previously optimized
pulses on one device serve as warm starts for synthesis on a
related device. Finally, integrating HyPulse with higher-level
hybrid compilers such as HybridLane~\cite{HybridLane2026} would
enable end-to-end compilation from algorithm-level circuits to
hardware-ready pulse schedules, completing the vertical stack
sketched in Fig.~\ref{fig:stack}.
\section*{Acknowledgments}
This work was supported in part by National Science Foundation grants
OSI-2531350, PHY-2325080a and OMA-2120757.
Yuan Liu and Frank Mueller were also supported in part by the
U.S. Department of Energy, Office of Science, Advanced Scientific
Computing Research, under contract number DE-SC0025384.

\bibliographystyle{IEEEtran}
\bibliography{qce2026}

@article{Gottesman2001,
  author  = {Gottesman, D. and Kitaev, A. and Preskill, J.},
  title   = {Encoding a qubit in an oscillator},
  journal = {Phys. Rev. A},
  volume  = {64},
  pages   = {012310},
  year    = {2001},
  doi     = {10.1103/PhysRevA.64.012310}
}

@article{Cirac1995,
  author  = {Cirac, J. I. and Zoller, P.},
  title   = {Quantum computations with cold trapped ions},
  journal = {Phys. Rev. Lett.},
  volume  = {74},
  pages   = {4091--4094},
  year    = {1995},
  doi     = {10.1103/PhysRevLett.74.4091}
}

@article{Crane2024HybridFermionBoson,
  author  = {Crane, Eleanor and Smith, Kevin C. and Tomesh, Teague
             and Eickbusch, Alec and Martyn, John M. and K{\"u}hn, Stefan
             and Funcke, Lena and DeMarco, Michael Austin and Chuang, Isaac L.
             and Wiebe, Nathan and Schuckert, Alexander and Girvin, Steven M.},
  title   = {Hybrid Oscillator-Qubit Quantum Processors: Simulating Fermions,
             Bosons, and Gauge Fields},
  journal = {arXiv preprint arXiv:2409.03747},
  year    = {2024}
}

@article{Khaneja2005,
  author  = {Khaneja, N. and Reiss, T. and Kehlet, C.
             and Schulte-Herbr{\"u}ggen, T. and Glaser, S. J.},
  title   = {Optimal control of coupled spin dynamics: Design of {NMR}
             pulse sequences by gradient ascent algorithms},
  journal = {J. Magn. Reson.},
  volume  = {172},
  pages   = {296--305},
  year    = {2005},
  doi     = {10.1016/j.jmr.2004.11.004}
}

@article{Leibfried2003,
  author  = {Leibfried, D. and Blatt, R. and Monroe, C. and Wineland, D.},
  title   = {Quantum dynamics of single trapped ions},
  journal = {Rev. Mod. Phys.},
  volume  = {75},
  pages   = {281},
  year    = {2003},
  doi     = {10.1103/RevModPhys.75.281}
}

@article{Liu2026PRX,
  author  = {Liu, Y. and Singh, S. and Smith, K. C. and others},
  title   = {Hybrid Oscillator--Qubit Quantum Processors: Instruction Set
             Architectures, Abstract Machine Models, and Applications},
  journal = {PRX Quantum},
  volume  = {7},
  pages   = {010201},
  year    = {2026},
  doi     = {10.1103/4rf7-9tfx}
}

@article{Lloyd1999,
  author  = {Lloyd, S. and Braunstein, S. L.},
  title   = {Quantum computation over continuous variables},
  journal = {Phys. Rev. Lett.},
  volume  = {82},
  pages   = {1784--1787},
  year    = {1999},
  doi     = {10.1103/PhysRevLett.82.1784}
}

@article{Montgomery2025HybridSimTI,
  author  = {Araz, J. Y. and Grau, M. and Montgomery, J. and Ringer, F.},
  title   = {Hybrid Quantum Simulations with Qubits and Qumodes
             on Trapped-Ion Platforms},
  journal = {Phys. Rev. A},
  volume  = {112},
  pages   = {012620},
  year    = {2025},
  doi     = {10.1103/kbv4-jj51}
}

@article{Weedbrook2012,
  author  = {Weedbrook, C. and Pirandola, S. and Garc\'{i}a-Patr\'on, R.
             and Cerf, N. J. and Ralph, T. C. and Shapiro, J. H. and Lloyd, S.},
  title   = {Gaussian quantum information},
  journal = {Rev. Mod. Phys.},
  volume  = {84},
  pages   = {621--669},
  year    = {2012},
  doi     = {10.1103/RevModPhys.84.621}
}

@article{Haffner2008,
  author  = {H{\"a}ffner, H. and Roos, C. F. and Blatt, R.},
  title   = {Quantum computing with trapped ions},
  journal = {Phys. Rep.},
  volume  = {469},
  pages   = {155--203},
  year    = {2008},
  doi     = {10.1016/j.physrep.2008.09.003}
}

@article{Andersen2015,
  author  = {Andersen, U. L. and Neergaard-Nielsen, J. S.
             and van Loock, P. and Furusawa, A.},
  title   = {Hybrid discrete- and continuous-variable quantum information},
  journal = {Nature Physics},
  volume  = {11},
  pages   = {713--719},
  year    = {2015},
  doi     = {10.1038/nphys3410}
}

@article{KatzCetinaMonroe2023PRXQ,
  author  = {Katz, O. and Cetina, M. and Monroe, C.},
  title   = {Programmable {$N$}-Body Interactions with Trapped Ions},
  journal = {PRX Quantum},
  volume  = {4},
  pages   = {030311},
  year    = {2023},
  doi     = {10.1103/PRXQuantum.4.030311}
}

@article{Matsos2025,
  author  = {Matsos, V. G. and Valahu, C. H. and Millican, M. J.
             and Navickas, T. and Kolesnikow, X. C.
             and Biercuk, M. J. and Tan, T. R.},
  title   = {Universal Quantum Gate Set for {Gottesman--Kitaev--Preskill}
             Logical Qubits},
  journal = {Nature Physics},
  year    = {2025},
  doi     = {10.1038/s41567-025-03002-8}
}

@article{Kienzler2015,
  author  = {Kienzler, D. and Lo, H.-Y. and Keitch, B.
             and de Clercq, L. and Leupold, F. and Lindenfelser, F.
             and Marinelli, M. and Negnevitsky, V. and Home, J. P.},
  title   = {Quantum harmonic oscillator state synthesis by reservoir engineering},
  journal = {Science},
  volume  = {347},
  pages   = {53--56},
  year    = {2015},
  doi     = {10.1126/science.1261033}
}

@article{Eickbusch2022,
  author  = {Eickbusch, A. and Sivak, V. and Ding, A. Z. and others},
  title   = {Fast universal control of an oscillator with weak dispersive
             coupling to a qubit},
  journal = {Nature Physics},
  volume  = {18},
  pages   = {1464--1469},
  year    = {2022},
  doi     = {10.1038/s41567-022-01776-9}
}

@article{Kang2021Duke,
  author  = {Kang, Mingyu and Liang, Qiyao and Zhang, Bichen
             and Huang, Shilin and Wang, Ye and Fang, Chao
             and Kim, Jungsang and Brown, Kenneth R.},
  title   = {Batch Optimization of Frequency-Modulated Pulses
             for Robust Two-Qubit Gates in Ion Chains},
  journal = {Phys. Rev. Applied},
  volume  = {16},
  number  = {2},
  pages   = {024039},
  year    = {2021},
  doi     = {10.1103/PhysRevApplied.16.024039}
}

@misc{QuTiP,
  author  = {Johansson, J. R. and Nation, P. D. and Nori, F.},
  title   = {{QuTiP}: An open-source {Python} framework for the
             dynamics of open quantum systems},
  journal = {Comput. Phys. Commun.},
  volume  = {183},
  pages   = {1760--1772},
  year    = {2012},
  doi     = {10.1016/j.cpc.2012.02.021}
}

@inproceedings{Liang2023HybridGatePulse,
  author    = {Liang, Zhiding and Song, Zhixin and Cheng, Jinglei
               and He, Zichang and Liu, Ji and Wang, Hanrui
               and Qin, Ruiyang and Wang, Yiru and Han, Song
               and Qian, Xuehai and Shi, Yiyu},
  title     = {Hybrid Gate-Pulse Model for Variational Quantum Algorithms},
  booktitle = {2023 60th ACM/IEEE Design Automation Conference (DAC)},
  year      = {2023},
  publisher = {IEEE},
  doi       = {10.1109/DAC56929.2023.10247923}
}

@inproceedings{Chen2023PAQOC,
  author    = {Chen, Yanhao and Jin, Yuwei and Hua, Fei and Hayes, Ari and Li, Ang and Shi, Yunong and Zhang, Eddy Z.},
  title     = {A Pulse Generation Framework with Augmented Program-aware Basis Gates and Criticality Analysis},
  booktitle = {2023 IEEE International Symposium on High-Performance Computer Architecture (HPCA)},
  year      = {2023},
  pages     = {773--786},
  publisher = {IEEE},
  doi       = {10.1109/HPCA56546.2023.10070990}
}

@inproceedings{EPOC2025,
  author    = {Cheng, Jinglei and Zhu, Yuchen and Zhou, Yidong
               and Ren, Hang and Song, Zhixin and Liang, Zhiding},
  title     = {{EPOC}: An Efficient Pulse Generation Framework with
               Advanced Synthesis for Quantum Circuits},
  booktitle = {2025 62nd ACM/IEEE Design Automation Conference (DAC)},
  year      = {2025},
  pages     = {1--7},
  publisher = {IEEE},
  doi       = {10.1109/DAC63849.2025.11133150}
}

@inproceedings{pulselib2024,
  author    = {Dalvi, Aniket S. and Riesebos, Leon and Whitlow, Jacob
               and Brown, Kenneth R.},
  title     = {Graph-Based Pulse Representation for Diverse Quantum
               Control Hardware},
  booktitle = {2024 IEEE International Conference on Quantum Computing
               and Engineering (QCE)},
  year      = {2024},
  pages     = {897--908},
  publisher = {IEEE},
  doi       = {10.1109/QCE60285.2024.00109}
}

@article{Wineland1998,
  author  = {Wineland, D. J. and Monroe, C. and Itano, W. M.
             and Leibfried, D. and King, B. E. and Meekhof, D. M.},
  title   = {Experimental issues in coherent quantum-state manipulation
             of trapped atomic ions},
  journal = {J. Res. Natl. Inst. Stand. Technol.},
  volume  = {103},
  pages   = {259--328},
  year    = {1998},
  doi     = {10.6028/jres.103.019}
}

@article{Stefanazzi2022,
  author  = {Stefanazzi, L. and Treptow, K. and Wilcer, N. and others},
  title   = {The {QICK} ({Quantum Instrumentation Control Kit}):
             Readout and control for qubits and detectors},
  journal = {Rev. Sci. Instrum.},
  volume  = {93},
  pages   = {044709},
  year    = {2022},
  doi     = {10.1063/5.0076249}
}

@article{Xu2021QuBIC,
  author  = {Xu, Y. and Huang, G. and Balewski, J. and others},
  title   = {{QubiC}: An open-source {FPGA}-based control and
             measurement system for superconducting quantum
             information processors},
  journal = {IEEE Trans. Quantum Eng.},
  volume  = {2},
  pages   = {1--11},
  year    = {2021},
  doi     = {10.1109/TQE.2021.3116883}
}

@article{Jaqal2020,
  author  = {Clark, S. M. and Lobser, D. and Revelle, M. C. and others},
  title   = {Engineering the {Quantum Scientific Computing Open
             User Testbed} ({QSCOUT}): Design details and user guide},
  journal = {arXiv preprint arXiv:2104.00759},
  year    = {2021}
}

@misc{ARTIQ,
  author       = {{M-Labs}},
  title        = {{ARTIQ}: Advanced Real-Time Infrastructure for
                 Quantum physics},
  howpublished = {\url{https://m-labs.hk/artiq}},
  year         = {2014}
}

@article{Mirrahimi2014,
  author  = {Mirrahimi, M. and Leghtas, Z. and Albert, V. V. and others},
  title   = {Dynamically protected cat-qubits: a new paradigm for
             universal quantum computation},
  journal = {New J. Phys.},
  volume  = {16},
  pages   = {045014},
  year    = {2014},
  doi     = {10.1088/1367-2630/16/4/045014}
}

@article{Grimsmo2020,
  author  = {Grimsmo, A. L. and Combes, J. and Baragiola, B. Q.},
  title   = {Quantum Computing with Rotation-Symmetric Bosonic Codes},
  journal = {PRX Quantum},
  volume  = {1},
  pages   = {010321},
  year    = {2020},
  doi     = {10.1103/PRXQuantum.1.010321}
}

@article{MacDonell2021,
  author  = {MacDonell, R. J. and Navickas, T. and Wecker, D. and others},
  title   = {Analog Quantum Simulation of Chemical Dynamics},
  journal = {Chem. Sci.},
  volume  = {12},
  pages   = {9794--9805},
  year    = {2021},
  doi     = {10.1039/D1SC02105C}
}

@article{HybridLane2026,
  author  = {Furches, J. and others},
  title   = {{Hybridlane}: A Software Development Kit for
             Hybrid Continuous-Discrete Variable Quantum Computing},
  journal = {arXiv preprint arXiv:2603.10919},
  year    = {2026}
}

@article{McKay2017,
  author  = {McKay, D. C. and Wood, C. J. and Sheldon, S.
             and Chow, J. M. and Gambetta, J. M.},
  title   = {Efficient {Z} gates for quantum computing},
  journal = {Phys. Rev. A},
  volume  = {96},
  pages   = {022330},
  year    = {2017},
  doi     = {10.1103/PhysRevA.96.022330}
}

@article{Alexander2020QiskitPulse,
  author  = {Alexander, T. and Kanazawa, N. and Egger, D. J.
             and others},
  title   = {Qiskit {Pulse}: Programming quantum computers through
             the cloud with pulses},
  journal = {Quantum Sci. Technol.},
  volume  = {5},
  pages   = {044006},
  year    = {2020},
  doi     = {10.1088/2058-9565/aba404}
}

@article{McKay2018OpenPulse,
  author  = {McKay, D. C. and Alexander, T. and Bello, L.
             and others},
  title   = {Qiskit backend specifications for {OpenQASM} and
             {OpenPulse} experiments},
  journal = {arXiv preprint arXiv:1809.03452},
  year    = {2018}
}

@misc{BoulderOpal,
  author       = {{Q-CTRL}},
  title        = {Boulder {Opal}: Quantum control infrastructure
                 software},
  howpublished = {\url{https://q-ctrl.com/boulder-opal}},
  year         = {2021}
}

@article{Sun2025SpinBoson,
  author  = {Sun, K. and Kang, M. and Nuomin, H. and Schwartz, G.
             and Beratan, D. N. and Brown, K. R. and Kim, J.},
  title   = {Quantum simulation of spin-boson models with
             structured bath},
  journal = {Nat. Commun.},
  volume  = {16},
  pages   = {4042},
  year    = {2025},
  doi     = {10.1038/s41467-025-59296-y}
}

@inproceedings{Riesebos2022DAX,
  author    = {Riesebos, Leon and Bondurant, Brad and Whitlow, Jacob and 
               Kim, Junki and Kuzyk, Mark and Chen, Tianyi and Phiri, Samuel and 
               Wang, Ye and Fang, Chao and Van Horn, Andrew and Kim, Jungsang and 
               Brown, Kenneth R.},
  title     = {Modular Software for Real-Time Quantum Control Systems},
  booktitle = {2022 IEEE International Conference on Quantum Computing and Engineering (QCE)},
  publisher = {IEEE},
  year      = {2022},
  pages     = {545--555},
  doi       = {10.1109/QCE53715.2022.00077}
}

\end{document}